\documentclass[12pt]{article}
\usepackage[margin=1.9cm]{geometry}
\usepackage{graphicx}
\usepackage{cite}
\usepackage{float}

\newcommand{\mysection}{\setcounter{equation}{0}\section}

\def\beq{\begin{equation}}
\def\eeq{\end{equation}}
\def\beqa{\begin{eqnarray}}
\def\eeqa{\end{eqnarray}}
 
\begin{document}

\begin{center}
{\Large \bf Higher-order soft-gluon corrections for $t{\bar t}Z$ cross sections}
\end{center}

\vspace{2mm}

\begin{center}
{\large Nikolaos Kidonakis and Chris Foster}\\

\vspace{2mm}

{\it Department of Physics, Kennesaw State University, \\
Kennesaw, GA 30144, USA}

\end{center}

\begin{abstract}
We calculate higher-order soft-gluon corrections for the associated production of a top-antitop quark pair and a $Z$ boson ($t{\bar t}Z$ production) in proton-proton collisions. We find that these corrections are numerically dominant and large. We present approximate NNLO (aNNLO) and approximate N$^3$LO (aN$^3$LO) cross sections that add, respectively, second-order and third-order soft-gluon corrections to the exact NLO QCD result. We also add electroweak (EW) corrections through NLO. We estimate uncertainties that arise from scale dependence, which are smaller at higher orders, and from parton distribution functions. We compare our aN$^3$LO QCD + NLO EW theoretical results to measurements of $t{\bar t}Z$ cross sections from the LHC and find good agreement. We also calculate single-particle-inclusive differential distributions in top-quark transverse momentum and rapidity.
\end{abstract}

\mysection{Introduction}

The production of a top-antitop quark pair in association with a $Z$ boson, i.e. $t{\bar t}Z$ production, is an interesting process that is crucial for measuring the coupling of the top quark to the $Z$ boson and relevant to the search for new physics, and which has been observed at the LHC. Cross sections of $t{\bar t}Z$ production have been measured at the LHC at 7 TeV \cite{CMS7}, 8 TeV \cite{CMS8a,ATLAS8,CMS8b}, and 13 TeV \cite{ATLAS13a,CMS13a,ATLAS13b,CMS13b,ATLAS13c,CMS13c,CMS13d,ATLAS13d} energies. Calculations at higher orders in the perturbative series are needed for better predictions of the rates with smaller uncertainties and, thus, more meaningful comparison to the data. 

The next-to-leading-order (NLO) QCD corrections for this process are large, and they were calculated in Refs. \cite{LMMP,KTP}. In addition, parton showers were included in \cite{GKPT1,GKPT2,MPT}, and electroweak corrections were computed in \cite{FHPSZ}. SMEFT studies for this process were presented in Refs. \cite{BI,AG}. The large size of the NLO corrections motivates the study of even higher orders.

An important and numerically dominant set of the contributions to the higher-order corrections for top-quark processes comes from the emission of soft, i.e. low energy, gluons near partonic threshold due to the large top-quark mass. We employ the soft-gluon resummation formalism of Refs. \cite{NKGS1,NKGS2,NKtt,NK2loop,NKtt2l} which has been used to study total and differential cross sections in top-antitop pair production \cite{NKtt,NKtt2l,NKttaN3LO,NKptyaN3LO,KGT} and single-top production \cite{NKsingletop,NKstLHC,NKsch,NKtW,NKtch,NKNY}. More recently, the resummation formalism was extended \cite{FK2020} to $2 \to 3$ processes in single-particle-inclusive (1PI) kinematics, in particular $tqH$ production \cite{FK2021}, $tq\gamma$ production \cite{NKNY2022}, $tqZ$ production \cite{NKNYtqZ}, $t{\bar t} \gamma$ production \cite{NKAT}, and $t{\bar t}W$ production \cite{KFttW}. In all these processes, as well as in $t{\bar t}Z$ production, the soft-gluon corrections account for the majority of the complete corrections at NLO and - where known - next-to-next-to-leading order (NNLO). 

In this paper, we calculate soft-gluon corrections for $t{\bar t}Z$ production by implementing the specific extension of the formalism presented in Ref. \cite{FK2020} and used in our previous paper on the related process of $t{\bar t}W$ production \cite{KFttW}. Alternative resummation formalisms for $t{\bar t}Z$ production, using different kinematics, i.e. the invariant mass of the $t{\bar t}Z$ system, were employed in Refs. \cite{KMSST,BFFPPT}. We stress that both the theoretical formalism/framework for the resummation and the choice of kinematics can significantly affect the numerical values of the results (see e.g. the review paper of Ref. \cite{NKBP} for a detailed discussion of various distinct formalisms and approaches). 

By expanding the resummed cross section to fixed order, we calculate approximate NNLO (aNNLO) and approximate next-to-next-to-next-to-leading-order (aN$^3$LO) cross sections and top-quark differential distributions for $t{\bar t}Z$ production. We derive the aNNLO results by adding the second-order soft-gluon corrections to the exact NLO calculation, and the aN$^3$LO results by further adding the third-order soft-gluon corrections to the aNNLO calculation. Moreover, we add electroweak (EW) corrections through NLO, thus deriving aN$^3$LO QCD + NLO EW results.

In the next section, we briefly discuss soft-gluon resummation for $t{\bar t}Z$ production. In Section 3, we present results for the total cross section at LHC energies through aN$^3$LO QCD + NLO EW. In Section 4, we present results for differential distributions in top-quark transverse momentum, $p_T$, and rapidity. We conclude in Section 5.

\mysection{Soft-gluon resummation for $t{\bar t}Z$ production}

We begin with a brief discussion of the formalism for soft-gluon resummation in $t{\bar t}Z$ production. At leading order (LO), the parton-level processes are $a(p_a)+b(p_b) \to t(p_t)+{\bar t}(p_{\bar t})+Z(p_Z)$, where $a$ and $b$ denote quarks and antiquarks or gluons, and we define the standard kinematical variables $s=(p_a+p_b)^2$, $t=(p_a-p_t)^2$, and $u=(p_b-p_t)^2$. We consider the emission of an additional gluon with momentum $p_g$ in the final state, and we define a threshold variable $s_4$ in 1PI kinematics, with the top quark being the observed particle with mass $m_t$, by $s_4=(p_{\bar t}+p_Z+p_g)^2-(p_{\bar t}+p_Z)^2=s+t+u-m_t^2-(p_{\bar t}+p_Z)^2$ which depends on the energy of the gluon and which vanishes when $p_g \to 0$. The soft-gluon corrections appear at each order in the perturbative series via logarithms of $s_4$, i.e. $\ln^k(s_4/m_t^2)/s_4$, where for the $n$th-order corrections the integer $k$ takes values from 0 through $2n-1$.

Soft-gluon resummation is derived from the factorization of the cross section under Laplace transforms as a product of functions that describe the hard scattering as well as collinear and soft emission, and the renormalization-group evolution of these functions \cite{NKGS1,NKGS2,NKtt,NKtt2l,FK2020,KFttW,GS,LOS}. More details are given in the related paper on $t{\bar t}W$ production \cite{KFttW} for which the resummation formalism is the same as in the present case. The resummed cross section can be expanded to fixed order, and the fixed-order results can then be inverted back to momentum space without need for a prescription \cite{NKtt,NKtt2l,NKttaN3LO,NKptyaN3LO,KGT,NKsingletop,NKstLHC,NKsch,NKtW,NKtch,NKNY,FK2020,FK2021,NKNY2022,NKNYtqZ,NKAT,KFttW}.

As discussed in \cite{KFttW}, when comparing resummation formalisms and frameworks at any logarithmic accuracy, in addition to noting the formalism (e.g. resummation in Mellin-moment space or under Laplace transforms or via soft-collinear effective theory), it is important to specify the kinematics and the variable used in the logarithms. In the 1PI kinematics that we use here, we have logarithms of $s_4/m_t^2$ in the perturbative series, while in previous work \cite{KMSST,BFFPPT} a variable involving the invariant mass of the $t{\bar t}Z$ system was used. The use of different variables can have a big numerical impact and it also affects the choice of central scale, which we naturally take in 1PI kinematics as the mass of the observed top quark. Thus, some calculations may have the same formal logarithmic accuracy but in different formalisms and variables and, thus, produce different numerical results. It is also important to specify if any prescriptions are used (and the associated ambiguities) or if fixed-order expansions are employed, as in this work. A good test for a formalism is its predictive power for exact fixed-order calculations. The formalism that we use here was the most successful \cite{NKtt2l} in predicting the NNLO cross section for $t{\bar t}$ production to very high accuracy and, as we showed recently in \cite{KFttW}, is consistent with the (partial) NNLO result in Ref. \cite{BDGKMRS} for the related process of $t{\bar t}W$ production.

\mysection{Total cross sections for $t{\bar t}Z$ production}

In this section, we present results for $t{\bar t}Z$ total cross sections. In our calculations, we set the factorization and renormalization scales equal to each other, and we denote this common scale by $\mu$. As mentioned previously, since we use 1PI kinematics in the soft-gluon resummation, we make the natural choice of using the mass of the observed top quark, $m_t=172.5$ GeV, as the central scale in our results below. 

The complete NLO results include both QCD and EW corrections, and they are calculated with {\small \sc MadGraph5\_aMC@NLO} \cite{MG5,MGew}. By adding the second-order soft-gluon corrections to the NLO QCD and NLO QCD+EW results, we derive aNNLO QCD and aNNLO QCD + NLO EW cross sections, respectively. By further adding the third-order soft-gluon corrections to the aNNLO QCD and aNNLO QCD + NLO EW results, we derive aN$^3$LO QCD and aN$^3$LO QCD + NLO EW cross sections, respectively.

\begin{figure}[htbp]
\begin{center}
\includegraphics[width=88mm]{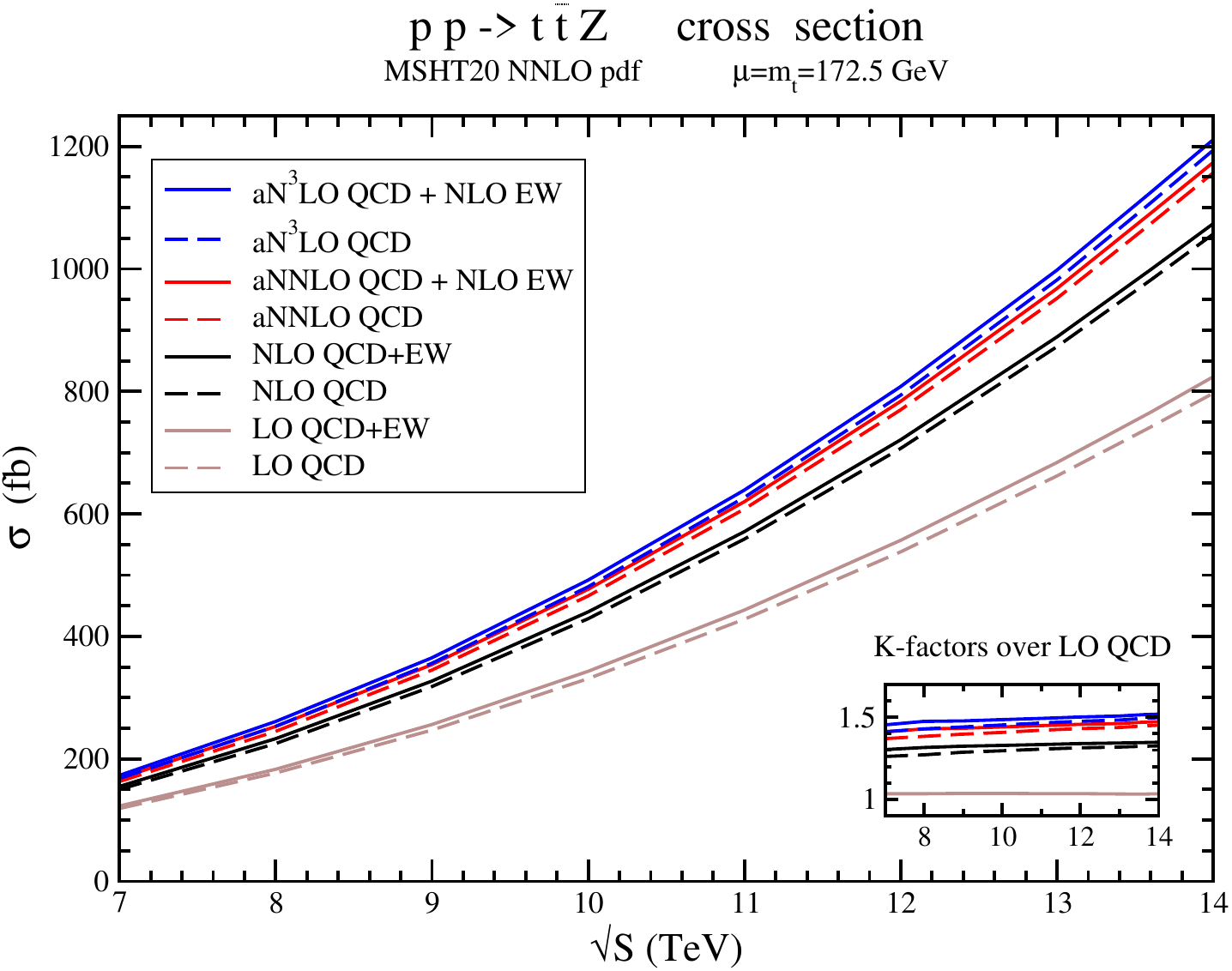}
\includegraphics[width=88mm]{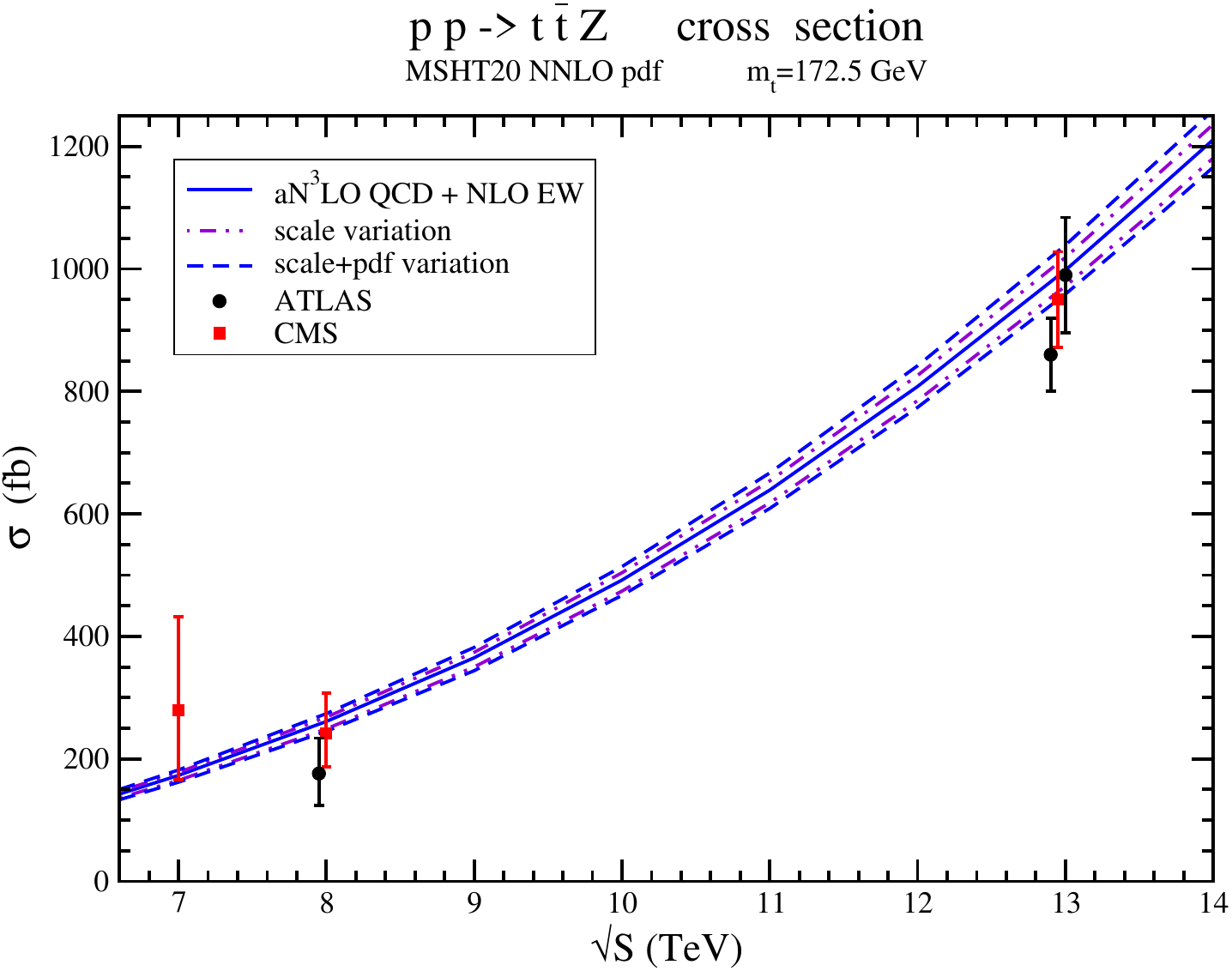}
\caption{The total cross sections for $t{\bar t}Z$ production in $pp$ collisions at LHC energies using MSHT20 NNLO pdf. The plot on the left shows cross sections at various orders from LO QCD through aN$^3$LO QCD + NLO EW, and the inset displays the $K$-factors of the higher orders relative to LO QCD. The plot on the right shows a comparison of the aN$^3$LO QCD + NLO EW cross section with LHC data at 7 TeV \cite{CMS7}, 8 TeV \cite{ATLAS8,CMS8b}, and 13 TeV \cite{CMS13b,ATLAS13c,ATLAS13d} energies.}
\label{ttZcs}
\end{center}
\end{figure}

In Fig. \ref{ttZcs}, we show results for the $t{\bar t}Z$ total cross section at LHC energies using MSHT20 NNLO pdf \cite{MSHT20NNLO}. The plot on the left shows results at four perturbative QCD orders, from LO to aN$^3$LO, both with and without electroweak corrections, for a total of eight combinations through aN$^3$LO QCD + NLO EW. The inset plot on the left shows the $K$-factors,  i.e. the ratios of the higher-order results relative to the LO QCD cross section. It is evident that the higher-order contributions are significant. The aN$^3$LO QCD + NLO EW $K$-factor is 1.45 at 7 TeV, 1.47 at 8 TeV, 1.51 at 13 TeV, and 1.52 at 13.6 and 14 TeV. 

The plot on the right in Fig. \ref{ttZcs} displays the aN$^3$LO QCD + NLO EW theoretical result with scale and pdf uncertainties, and compares it to LHC data from CMS at 7 TeV \cite{CMS7}, from ATLAS \cite{ATLAS8} and CMS \cite{CMS8b} at 8 TeV, and from CMS \cite{CMS13b} and ATLAS \cite{ATLAS13c,ATLAS13d} at 13 TeV. The scale uncertainties are obtained by varying the scale by a factor of two around the central choice, i.e. $m_t/2 \le \mu \le 2m_t$ while the pdf uncertainties are as provided by MSHT20 NNLO pdf. We observe a good agreement between the data and the theoretical calculation at all energies.

\begin{table}[H]
\begin{center}
\begin{tabular}{|c|c|c|c|c|c|c|c|c|} \hline
\multicolumn{6}{|c|}{$t{\bar t} Z$ cross sections in $pp$ collisions at the LHC} \\ \hline
$\sigma$ in fb & 7 TeV & 8 TeV & 13 TeV & 13.6 TeV & 14 TeV \\ \hline
LO QCD                & $119^{+44}_{-30}$ & $177^{+64}_{-44}$ & $662^{+218}_{-152}$ & $742^{+242}_{-169}$ & $ 797^{+258}_{-181}$ \\ \hline
LO QCD+EW             & $123^{+46}_{-31}$ & $183^{+67}_{-45}$ & $684^{+226}_{-157}$ & $766^{+250}_{-175}$ & $ 824^{+267}_{-187}$ \\ \hline
NLO QCD               & $150^{+ 9}_{-16}$ & $225^{+15}_{-24}$ & $873^{+ 69}_{- 93}$ & $982^{+ 79}_{-104}$ & $1057^{+ 85}_{-112}$ \\ \hline
NLO QCD+EW            & $155^{+ 8}_{-16}$ & $233^{+13}_{-24}$ & $889^{+ 61}_{- 90}$ & $999^{+ 70}_{-101}$ & $1074^{+ 74}_{-107}$ \\ \hline
aNNLO QCD             & $163^{+7}_{-10}$ & $245^{+10}_{-15}$ & $952^{+29}_{-48}$ & $1074^{+33}_{-54}$ & $1157^{+35}_{-58}$ \\ \hline
aNNLO QCD + NLO EW    & $168^{+6}_{-10}$ & $253^{+9}_{-15}$ & $968^{+25}_{-46}$ & $1091^{+29}_{-51}$ & $1174^{+30}_{-54}$ \\ \hline
aN$^3$LO QCD          & $168^{+5}_{-8}$ & $253^{+8}_{-12}$ & $982^{+25}_{-28}$ & $1108^{+28}_{-32}$ & $1194^{+30}_{-34}$ \\ \hline
aN$^3$LO QCD + NLO EW & $173^{+5}_{-8}$ & $261^{+7}_{-12}$ & $998^{+21}_{-26}$ & $1125^{+24}_{-30}$ & $1211^{+25}_{-30}$ \\ \hline
\end{tabular}
\caption[]{The $t{\bar t}Z$ cross sections (in fb) at various orders through aN$^3$LO QCD + NLO EW with scale uncertainties, in $pp$ collisions with $\sqrt{S}=7$, 8, 13, 13.6, and 14 TeV, $m_t=172.5$ GeV, and MSHT20 NNLO pdf.}
\label{table1}
\end{center}
\end{table}

In Table \ref{table1}, we provide numerical results for LO QCD through aN$^3$LO QCD + NLO EW total cross sections in $t{\bar t}Z$ production at 7, 8, 13, 13.6, and 14 TeV proton-proton collision energies using MSHT20 NNLO pdf \cite{MSHT20NNLO}. The results shown are with central scale $\mu=m_t$, and the uncertainties provided are for the usual scale variation by a factor of two. As expected, the scale uncertainties are significantly reduced at higher orders. At 13.6 TeV LHC energy and using the central scale, we see from the numbers in Table \ref{table1} that the NLO QCD corrections increase the LO QCD result by 32\%, the aNNLO corrections add another 12\%, and the aN$^3$LO corrections contribute an extra 5\%. The electroweak corrections through NLO provide an additional 2\%. Thus, at 13.6 TeV, the aN$^3$LO QCD + NLO EW cross section is 52\% larger than that at LO QCD.

\begin{table}[H]
\begin{center}
\begin{tabular}{|c|c|c|c|c|c|c|c|c|} \hline
\multicolumn{6}{|c|}{aN$^3$LO QCD + NLO EW $t{\bar t} Z$ cross section in $pp$ collisions at the LHC} \\ \hline
$\sigma$ in fb & 7 TeV & 8 TeV & 13 TeV & 13.6 TeV & 14 TeV \\ \hline
MSHT20 NNLO pdf & $173^{+5}_{-8}{}^{+4}_{-3}$ & $261^{+7}_{-12}{}^{+6}_{-4}$ & $998^{+21}_{-26}{}^{+20}_{-13}$ & $1125^{+24}_{-30}{}^{+22}_{-14}$ & $1211^{+25}_{-30}{}^{+23}_{-15}$ \\ \hline
MSHT20 aN$^3$LO pdf & $170^{+5}_{-8}{}^{+4}_{-4}$ & $255^{+7}_{-12}{}^{+6}_{-6}$ & $974^{+20}_{-25}{}^{+19}_{-21}$ & $1096^{+23}_{-29}{}^{+22}_{-23}$ & $1182^{+24}_{-29}{}^{+23}_{-25}$ \\ \hline
NNPDF4.0 aN$^3$LO pdf & $165^{+5}_{-7}{}^{+2}_{-2}$& $248^{+7}_{-11}{}^{+3}_{-3}$ & $962^{+20}_{-25}{}^{+6}_{-6}$ & $1083^{+23}_{-29}{}^{+7}_{-7}$ & $1170^{+24}_{-29}{}^{+7}_{-7}$ \\ \hline
\end{tabular}
\caption[]{The $t{\bar t}Z$ cross sections (in fb) at aN$^3$LO QCD + NLO EW with scale and pdf uncertainties, in $pp$ collisions with $\sqrt{S}=7$, 8, 13, 13.6, and 14 TeV, $m_t=172.5$ GeV, and MSHT20 NNLO pdf, MSHT20 aN$^3$LO pdf, and NNPDF4.0 aN$^3$LO pdf.}
\label{table2}
\end{center}
\end{table}

Recently, aN$^3$LO pdf sets have become available from MSHT \cite{MSHT20aN3LO} and NNPDF \cite{NNPDF40aN3LO}. While these sets are approximate, it makes sense to do calculations at aN$^3$LO QCD with those sets as well and compare the results. In Table \ref{table2}, we provide numerical results for the aN$^3$LO QCD + NLO EW total cross sections in $t{\bar t}Z$ production, including both scale and pdf uncertainties, using  MSHT20 NNLO pdf \cite{MSHT20NNLO}, MSHT20 aN$^3$LO pdf \cite{MSHT20aN3LO}, and NNPDF4.0 aN$^3$LO pdf \cite{NNPDF40aN3LO}. The MSHT20 aN$^3$LO pdf give cross sections a couple of percent or so smaller than those with MSHT20 NNLO pdf, and the cross sections with NNPDF4.0 aN$^3$LO pdf are a bit smaller yet.

Next, we provide some more details about the comparison between the LHC data at 7, 8, and 13 TeV and our best prediction. CMS measured a cross section of $0.28^{+0.14}_{-0.11}{}^{+0.06}_{-0.03}$ pb at 7 TeV \cite{CMS7}. At 8 TeV energy, the cross section measured by ATLAS is $176^{+58}_{-52}$ fb \cite{ATLAS8} and by CMS $242^{+65}_{-55}$ fb \cite{CMS8b}. More recently, CMS has measured a $t{\bar t}Z$ production cross section at 13 TeV of $0.95 \pm 0.05 \pm 0.06$ pb \cite{CMS13b}. ATLAS has measured a cross section at 13 TeV of $0.99 \pm 0.05 \pm 0.08$ pb in \cite{ATLAS13c} and $0.86 \pm 0.04 \pm 0.04$ pb in \cite{ATLAS13d}. It is clear that the experimental uncertainties are very big at 7 TeV and quite significant at 8 TeV, but they are much smaller at 13 TeV. The measured cross sections are in good agreement with our results that include the higher-order corrections as seen from Fig. \ref{ttZcs} and the numbers in Tables \ref{table1} and \ref{table2}. 

\mysection{Top-quark $p_T$ and rapidity distributions}

In this section we present results for the top-quark $p_T$ and rapidity distributions in $t{\bar t}Z$ production. 

\begin{figure}[htbp]
\begin{center}
\includegraphics[width=88mm]{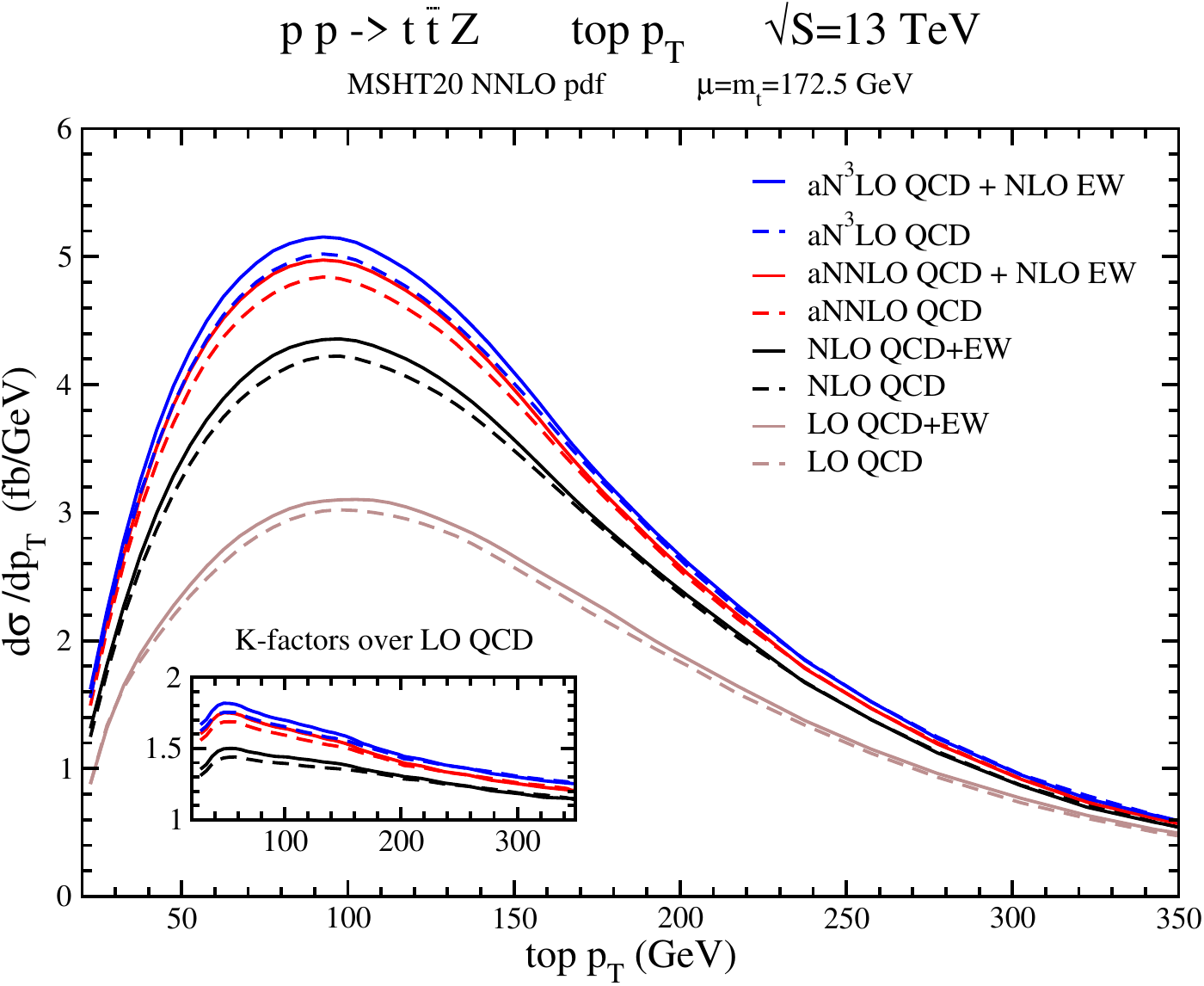}
\includegraphics[width=88mm]{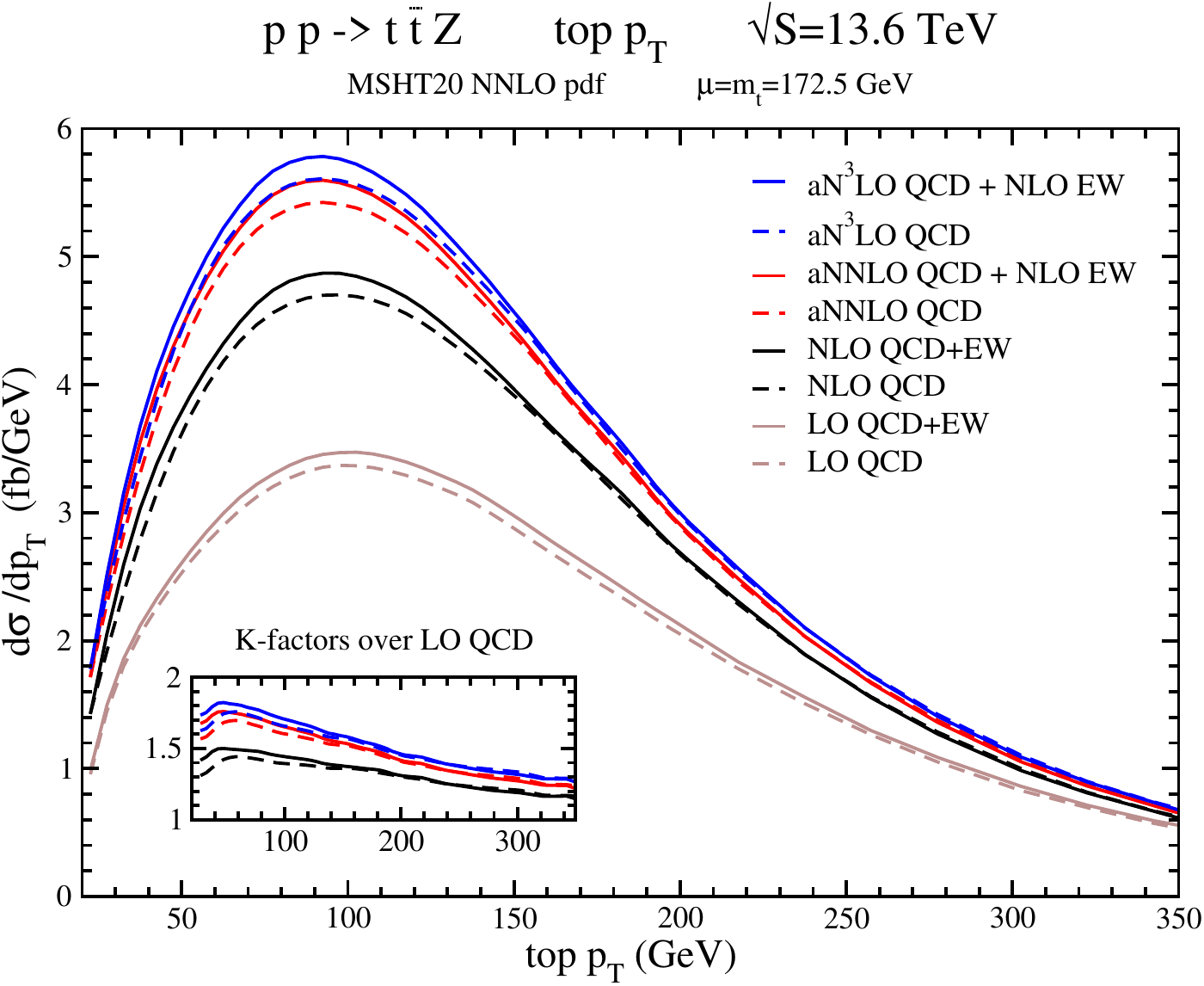}
\caption{The top-quark $p_T$ distributions at various orders through aN$^3$LO QCD + NLO EW in $t{\bar t}Z$ production in $pp$ collisions at LHC energies of 13 TeV (left plot) and 13.6 TeV (right plot) using MSHT20 NNLO pdf. The inset plots display the $K$-factors relative to LO QCD.}
\label{pttop}
\end{center}
\end{figure}

We begin with the $p_T$ distributions for recent and current LHC energies. In Fig. \ref{pttop} we plot the top-quark $p_T$ distributions at (left) 13 TeV and (right) 13.6 TeV energies using MSHT20 NNLO pdf. Results are shown at LO, NLO, aNNLO, and aN$^3$LO QCD both with and without the electroweak corrections. At both energies, the distributions peak at $p_T$ values between 90 and 100 GeV. The inset plots display the $K$-factors of the higher-order results relative to the LO QCD distributions. We note that the $K$-factors peak at relatively small values of $p_T$, around 50 GeV, and the higher-order contributions are very significant, giving as high as 82\% enhancement over LO QCD.

\begin{figure}[htbp]
\begin{center}
\includegraphics[width=88mm]{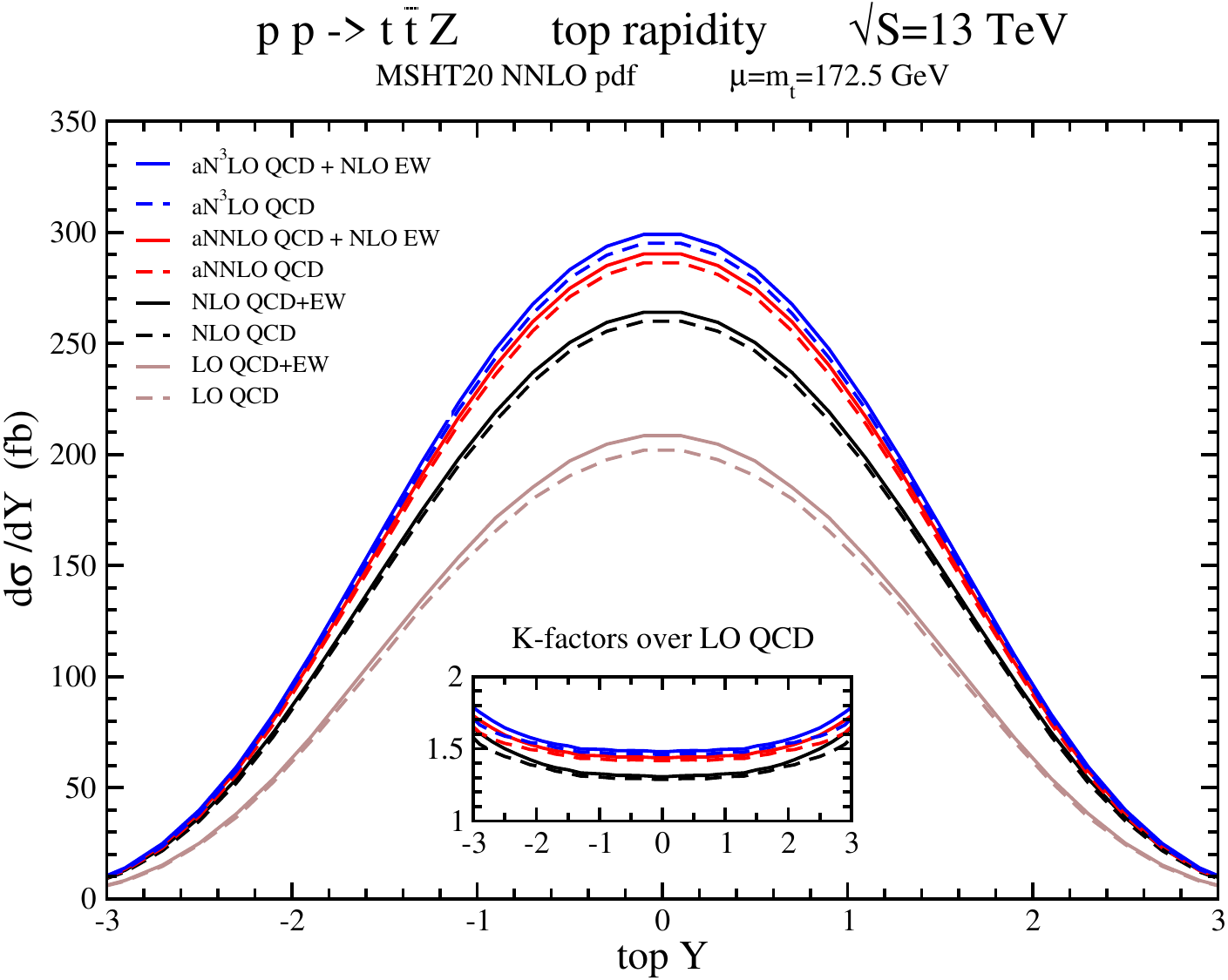}
\includegraphics[width=88mm]{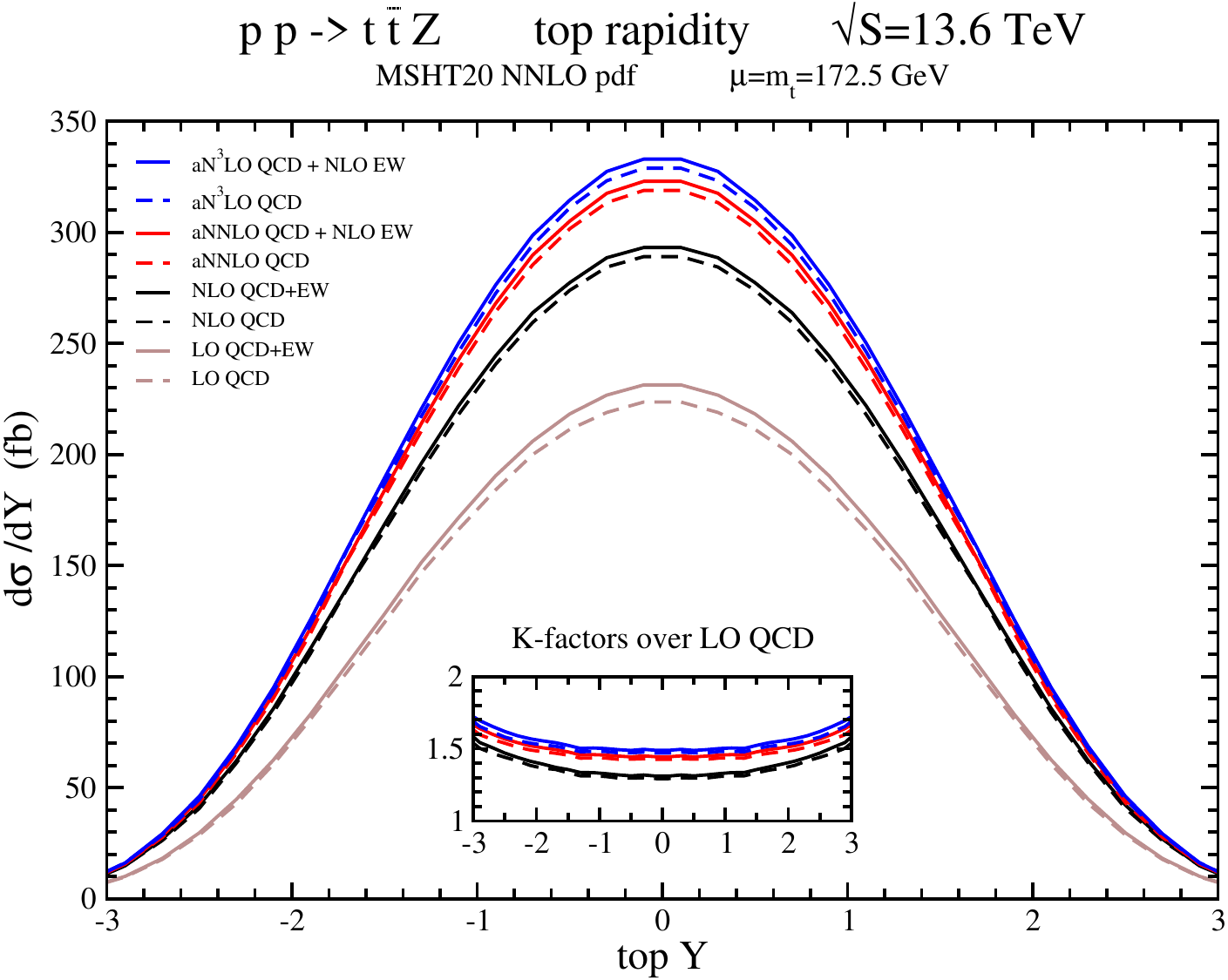}
\caption{The top-quark rapidity distributions at various orders through aN$^3$LO QCD + NLO EW in $t{\bar t}Z$ production in $pp$ collisions at LHC energies of 13 TeV (left plot) and 13.6 TeV (right plot) using MSHT20 NNLO pdf. The inset plots display the $K$-factors relative to LO QCD.}
\label{ytop}
\end{center}
\end{figure}

We continue with the rapidity distributions. In Fig. \ref{ytop} we plot the top-quark rapidity distributions at (left) 13 TeV and (right) 13.6 TeV energies using MSHT20 NNLO pdf. Again, results are shown at LO, NLO, aNNLO, and aN$^3$LO QCD both with and without the electroweak corrections. The inset plots display the $K$-factors relative to the LO QCD distributions, and we note that the higher-order contributions are very significant, particularly at larger rapidities. While the aN$^3$LO QCD+NLO EW over LO QCD $K$-factors are around 1.5 at central rapidities for both energies, they reach the value of 1.8 for a rapidity of 3 at 13 TeV energy.

There have not yet been measurements of the top-quark $p_T$ and rapidity distributions in $t{\bar t}Z$ production at the LHC, so it is not possible to make comparisons to data at this time. However, our theoretical predictions at higher orders significantly enhance the NLO distributions and, similar to the case of the total cross sectiosn, we can reasonably expect an improved description of future data.

\mysection{Conclusions}

We have presented theoretical predictions for total cross sections and top-quark differential distributions for $t{\bar t}Z$ production. The higher-order corrections are large and are dominated by contributions from soft-gluon emission. We have calculated second-order and third-order soft-gluon corrections and added them to the NLO QCD result. This is the first calculation of soft-gluon contributions for this process in single-particle-inclusive kinematics. The higher-order corrections decrease the dependence on the factorization and renormalization scales. We have further added NLO electroweak corrections to produce aN$^3$LO QCD + NLO EW predictions. The total cross sections are in good agreement with data from the LHC. We have also calculated the top-quark $p_T$ and rapidity distributions in $t{\bar t}Z$ production, again finding large contributions from the higher orders.

\section*{Acknowledgements}
This material is based upon work supported by the National Science Foundation under Grant No. PHY 2412071.

\end{document}